\documentclass[9pt,twocolumn,twoside]{osajnl}

\journal{ol} 

\setboolean{shortarticle}{true}

\title{Addressing asymmetric Fano profiles on molecular lines in dual-comb spectroscopy}

\author[1]{Philippe Guay}
\author[1]{Mathieu Walsh}
\author[1,*]{Jérôme Genest}

\affil[1]{Centre d'optique, photonique et laser, Universit\'{e} Laval, Qu\'{e}bec, Qu\'{e}bec G1V 0A6, Canada}
\affil[*]{Corresponding author: jerome.genest@copl.ulaval.com}

\begin{abstract}
Fano resonance in molecular spectroscopy is reported with a dual-comb instrument. The effect is observed as asymmetric absorption lines of H$^{12}$CN. Pulse chirping conditions in the gas cell are varied to show that Fano resonance is dependent on the pulse peak power. A model adding the Fano profile to Voigt lines is used to estimate Fano phase as a function of pulse peak power. A pulse peak power condition is derived from this analysis to avoid lineshape distortion in pulsed laser experiments. 
\end{abstract}

\begin{document}

\maketitle

Dual-comb spectroscopy's (DCS) high sensitivity is enabling improvements to theoretical models beyond the Voigt profile \cite{YAN18,COL21b}. For instance, speed-dependent collisions \cite{SCHRO18,COL21a} are now included in advanced spectroscopic models \cite{TEN14} and other non-Voigt effects such as line-mixing are studied with DCS \cite{COL21b}. Advanced spectroscopic models have now become essential to quantify atmospheric gases to the required part per billion level \cite{ADL10,HER21}, especially in a climate change context.  

Further improvements to the spectroscopic models with frequency combs are however hampered by systematic errors of comparable level. For instance, polarization effects have been reported to create spectral asymmetry \cite{NIS16}. Comb-based Fourier transform spectroscopy deals with an instrument lineshape that alters the absorption lineshapes if not taken into account \cite{MAS16,RUT18}. Cavity-enhanced dual-comb spectrometers suffer from asymmetric lineshape when the comb is detuned with respect to the cavity modes \cite{FOL13,FLE18}.  Increased power on photodetectors to improve the signal-to-noise ratio (SNR) has lead to their saturation and to systematic errors such as an overestimation of lines intensities \cite{GUA21b}. Towards this reach of better SNR, it has become judicious to use all the available comb signal and thus send high power pulses into the gas cell. This leads to systematic errors introducing line asymmetry and shift. These have been attributed to the Fano effect when probing microresonators \cite{HEB16,LI11} and further inspection of data presented in \cite{ROY12} reveals that molecular lines can also be affected, but this has never been properly addressed as a systematic error in dual-comb spectroscopy. 

In the strong field and attosecond pulses community, it is well known that short pulse excitation can couple a sharp resonance and an absorption continuum \cite{FAN61,BAN04,HEL81}. As a result, asymmetric absorption lines have been observed and quantified by a Fano profile \cite{OTT13}.

In this letter, a dual-comb spectrometer is used to show that excitation of gas molecules with sub-picosecond pulses may lead to asymmetric absorption lines. These lines are modeled with Fano profiles and their asymmetry is found to be dependent on the peak power of the pulses in the gas cell. Three gases interrogated at separate times with different spectrometers exhibit similar asymmetric Fano profile. To understand  the conditions under which this asymmetry arises, the peak power of the pulses interacting with H$^{12}$CN gas molecules is varied with different chirping conditions. The Fano asymmetry can be properly accounted by a modification of the Voigt profile \cite{OTT13} and is quantified by a parameter called the Fano phase. A plot of the Fano phase as a function of pulse peak power is presented for H$^{12}$CN around 1550 nm. 

This systematic effect that is a priori present in all DCS measurements leading to residuals on the order of a few percent around narrow spectral features can be addressed by including the proper Fano phase in the retrieval algorithms or by moderately chirping the pulses interrogating the sample to ten picoseconds. Such chirp can also be useful to optimize the SNR in the context of limited dynamic range \cite{GUA22a,ROY12}.

\begin{figure}[h]
\centering\includegraphics[width=0.5\textwidth]{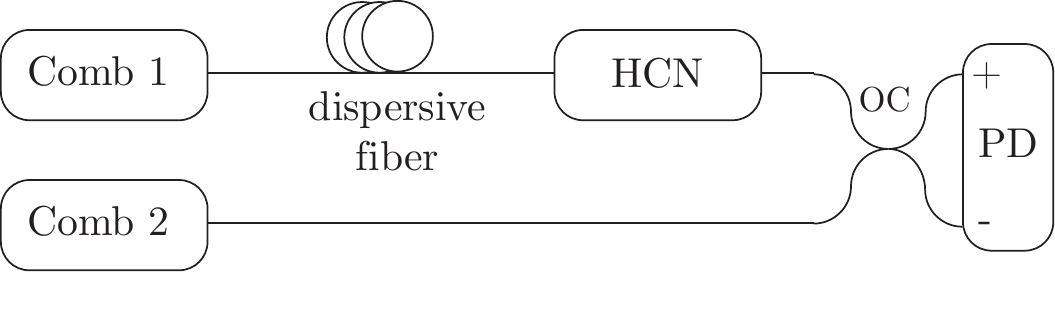}
\caption{Experimental setup where the gas probing pulses are stretched or compressed by dispersive fiber. HCN: Hydrogen cyanide gas cell. OC: 50/50 optical coupler. PD: Balanced photodetector.}
\label{setup}
\end{figure}

The experimental setup is shown in Fig. \ref{setup} where two mode-locked lasers based on the design in \cite{SIN15} are used. These saturable absorbant mirror (SAM) lasers operate at 1550 nm with a repetition rate of 160 MHz and a repetition rate difference of 1 kHz. The lasers output each 45 mW of average power. One laser is sent to optical fibers which are used to compress or stretch the pulses before entering the hydrogen cyanide gas cell (Wavelength References). Various length of single-mode fiber (SMF) with normal dispersion and inverse dispersion fibers (IDF) are used to vary the pulse peak power and duration in the gas cell.  The combs are mixed after probing the gas and dual-comb inteferograms are measured (BDX1BA from Thorlabs) and digitized (Gage CSE8389). 


Linear interferograms are measured using 30 mW of comb power as in \cite{GUA22a} and are filtered in a 60 MHz bandwidth to avoid aliasing in the acquisition card. Once digitized, the interferograms are phase-corrected using an acquired reference signal at 1549 nm (RIO Planex) and then a self-correction algorithm is applied \cite{HEB17}. The averaged interferogram is then Fourier transformed to yield the transmission spectrum of the probed gas. An HITRAN fit based on the Voigt profile is first performed on the observed lines. An optimization procedure is used to determine the cell's pressure, length and temperature. Limits are imposed on the cell's parameters to match the manufacturer's specifications and uncertainties. The optimization procedure uses the optical point spacing, the initial frequency and the coefficients of an 8th order polynomial as free parameters. The polynomial is used as a baseline calibration. 

The pulse chirp is varied by changing the fiber length in the sample arm (comb 1) while maintaining a nominal length in the local oscillator (LO) arm (comb 2). The pulse duration in the LO arm is measured to 4 ps using an autocorrelator. The interferogram phase is then used to retrieve the pulse duration in the sample arm for each fiber length. The sample arm pulse duration is expressed as the width of an equivalent square pulse, a metric similar to the noise equivalent bandwidth (NEB).

Although the pulse width is always a positive scalar, it is presented thereafter as positive and negative values to account for the normal and anomalous dispersion conditions. An asymmetric chirped pulse may have a different shape whether it is normally or abnormally dispersed, and thus may generate a different coupling between gas resonances. This effect is taken into account by distinguishing the normal and abnormal dispersion regime respectively by positive and negative pulse width. 

To determine the pulse peak power, the average power of the lasers is usually divided by the laser's repetition rate  and by the pulse width. In a setup where average power is constant throughout measurements, the pulse width is inversely proportional to the peak power. In this experiment however, the average power is not constant between measurements since many fibers with different losses are  combined to have distinctive dispersive coefficients. Moreover, the inverse dispersion fiber has a small core radius which introduces more losses than with standard single-mode fiber. As a result, the power is measured for each set of dispersive fibers and then normalized by the laser's repetition rate and the pulse width to yield the pulse peak power.



\begin{figure}[h]
\centering\includegraphics[width=0.5\textwidth]{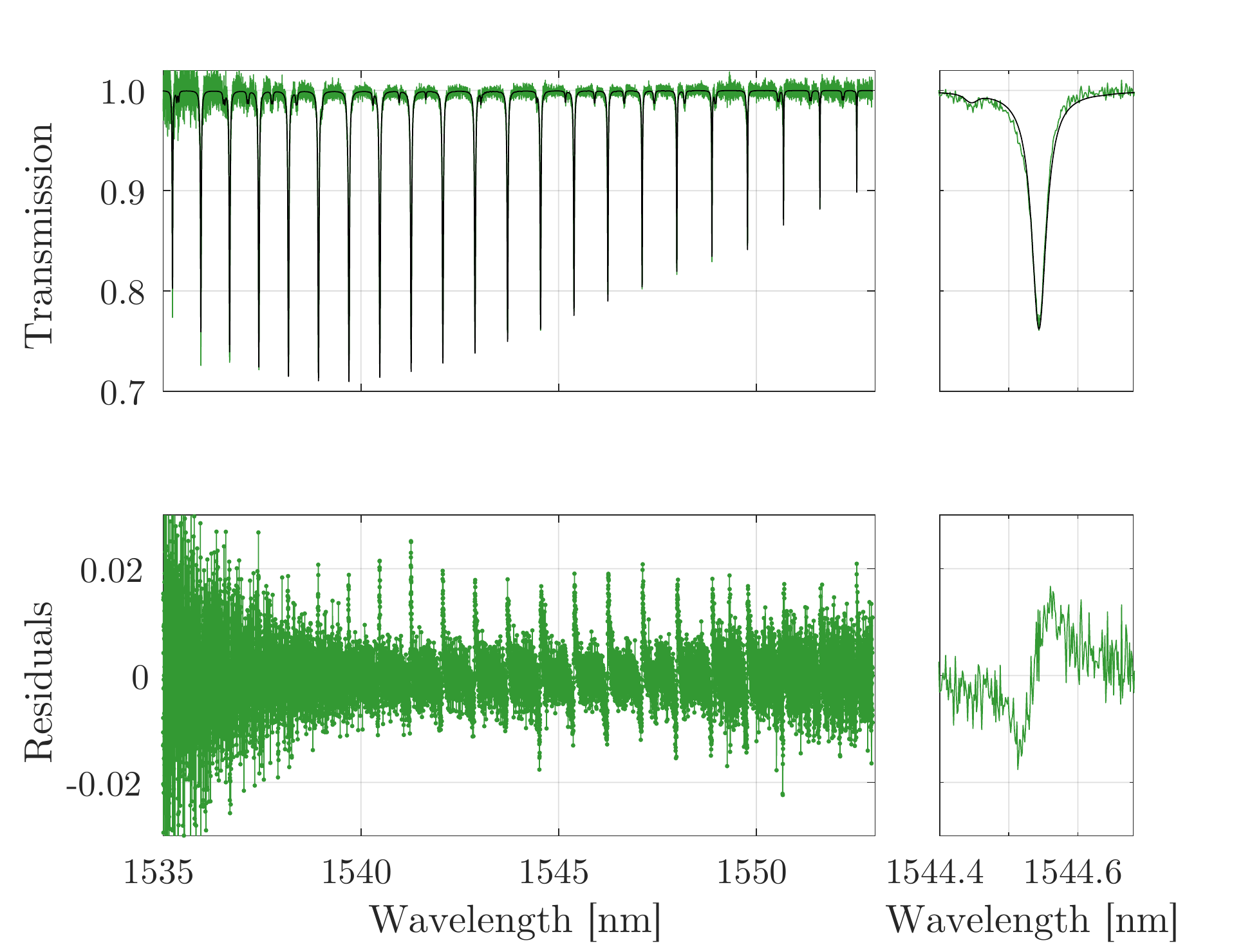}
\caption{(top left panel) Broadband transmission spectrum of H$^{12}$CN and (top right panel) zoom-in on a single absorption line. The residuals between the experimental data (green) and the HITRAN fit (black) are shown in the bottom panels.}
\label{H12CN_spec}
\end{figure}

A broadband transmission spectrum is shown on the top panel of Fig. \ref{H12CN_spec} for  pulses compressed to nearly 2 ps. An inverse dispersion fiber of 3 m and a single-mode fiber of 1 m have been added as dispersive fibers. This case is chosen to represent a typical measurement where no special dispersion management is implemented, but some non-negligible fiber length is present in the sample arm. The experimental data is shown in Fig. \ref{H12CN_spec} as well as an HITRAN fit based on the Voigt model. The residuals on the bottom plot of Fig. \ref{H12CN_spec} clearly show a dominating systematic error on the lines. The spikes in the residuals result from the experimental lineshapes' asymmetry. The right-hand side plots of Fig. \ref{H12CN_spec} focus on the 1544.55 nm line to explicitly show that right shoulder of the lines is pushed upwards while the left should is dragged downward, resulting in a skewed absorption.  The lower optical SNR at low and high wavelength explains the increased noise trace in the residuals. 

\begin{figure}[h]
\centering\includegraphics[width=0.5\textwidth]{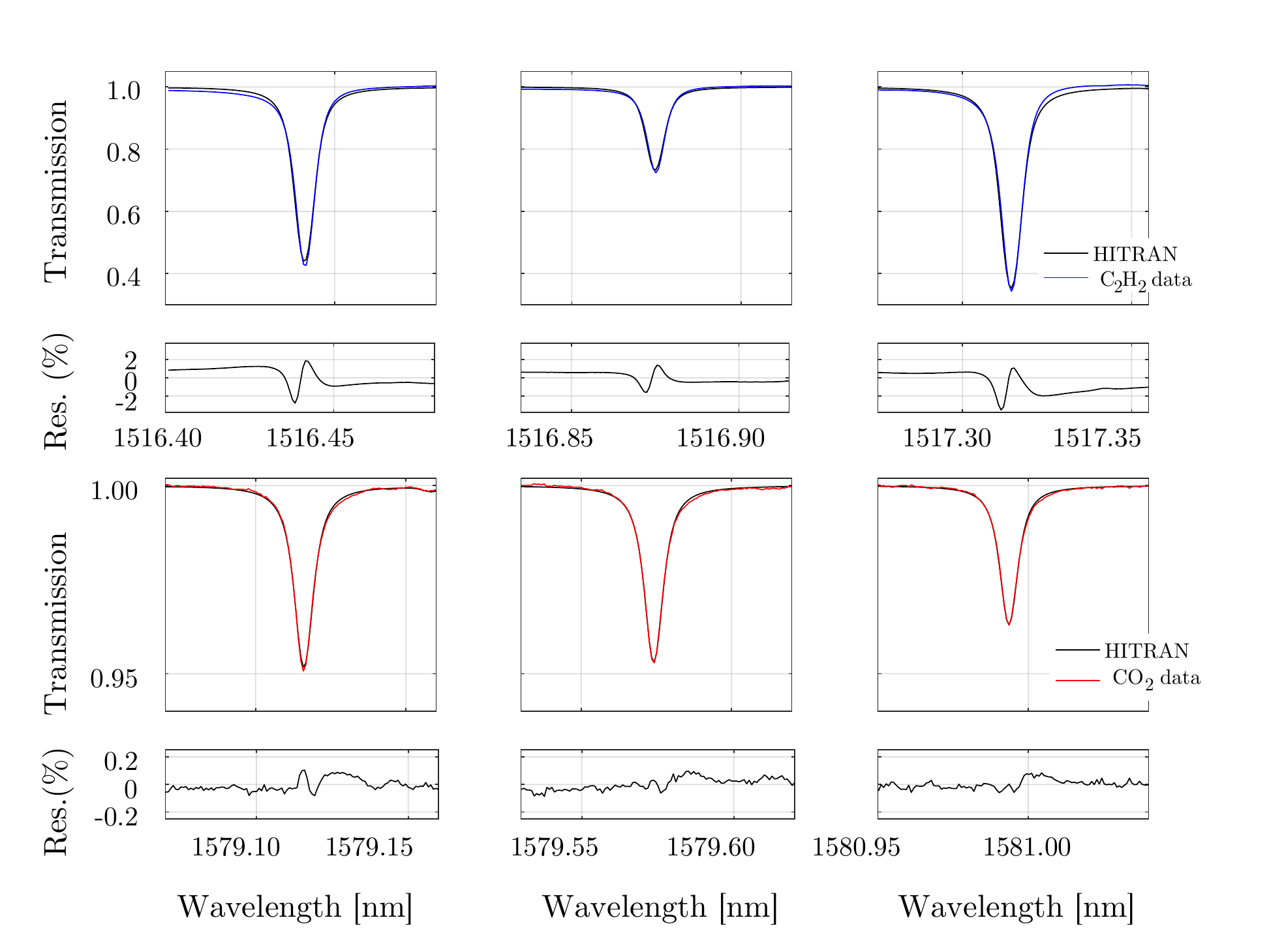}
\caption{(top panels) Transmission spectrum of C$_2$H$_2$ from \cite{ROY12} (blue) showing skewed absorption lines and the corresponding HITRAN fit (black). The residuals are shown as underneath panels. (middle panels) Transmission spectrum of CO$_2$ (red) with similar asymmetry and the corresponding HITRAN fit (black). The residuals are shown as underneath panels.}
\label{C2H2_CO2_spec}
\end{figure}

Asymmetric lines have been observed in previously published data as well. Further processing on the data in \cite{ROY12} has shown similar skewed lines with a different spectrometer and a different gas. The experiment was performed with Menlo c-combs with a repetition rate of 100 MHz. A mixture of C$_2$H$_{2}$ and H$^{13}$CN was measured. A few lines of the transmission spectrum of C$_2$H$_{2}$ are shown in the top  panels of Fig. \ref{C2H2_CO2_spec}. Three arbitrary lines are shown as well as the corresponding HITRAN fits. Each line shows a shoulder above the adjusted fit and another below. This effect is accentuated in the panels underneath the lines where the residuals are once again dominated by an asymmetric systematic error. 

Furthermore, a more recent dual-comb experiment with CO$_2$ performed with the same Menlo combs shares the same issue. A few lines are shown in the middle panels of Fig. \ref{C2H2_CO2_spec} (red curves). Similarly as the C$_2$H$_{2}$ experiment, the residuals show a systematic error, but it is more subtle. One has to look in the residuals, especially at the levels directly on the right-hand side of the line to show that it is systematically positive and on the left-hand side of the lines to find that it is systematically negative. 


The asymmetry of the lines in the examples above may differs from one another as the conditions of the experiments were significantly different. The pulse peak power in the gas sample was different as the pulse length and the average power were different. The length of dispersive fibers before the gas cell were not controlled and thus differ across measurements, leading to different peak power in the gas cell. Moreover, one has to consider the relative depth of the absorption lines to assess the level of asymmetry across gases since the residuals are an absolute measurement and not a relative one. For instance, C$_2$H$_2$ lines appear more asymmetric than CO$_2$ in Fig. \ref{C2H2_CO2_spec}, but the relative depth has to be taken into account as CO$_2$ lines depth are a tenth of the C$_2$H$_2$'s. Finally, the coupling leading to Fano asymmetry may also be stronger or weaker for different molecules.

\begin{figure}[!ht]
\centering\includegraphics[width=0.5\textwidth]{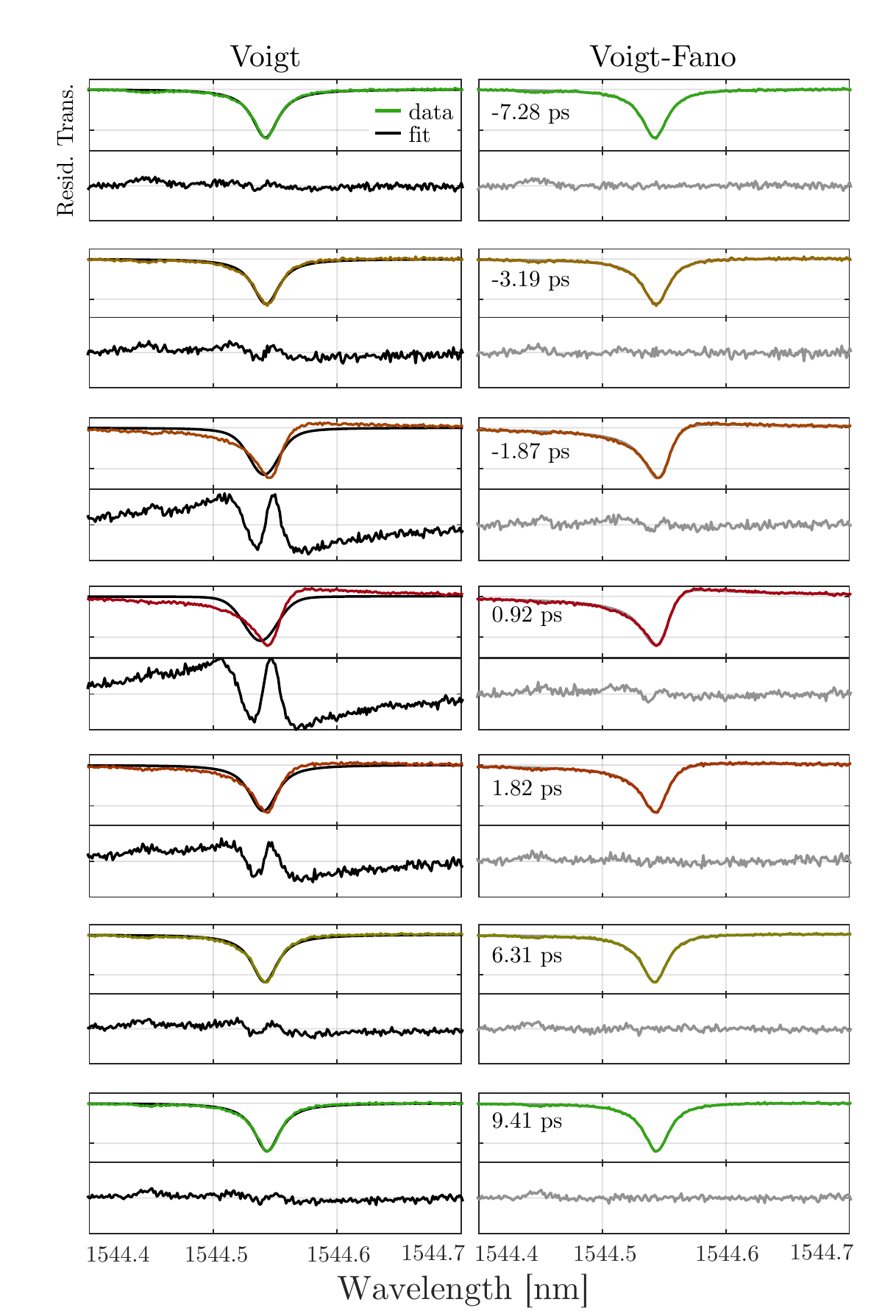}        
\caption{(top panel of subplots) Transmission spectrum of the 1544.55 nm line of H$^{12}$CN for experimental data (colors) and theoretical model (gray/black). (bottom panel of subplots) Residuals from the data and the model. The left column shows experimental data with a Voigt fit while the right column presents a Voigt-Fano fit.  From top to bottom, the panels show decreasing pulse width in the gas cell (green to red) and increasing pulse width (red to green).}
\label{Fano_lines}
\end{figure}

By varying the pulse duration and pulse peak power in a gas cell, it is possible to control the line asymmetry. A first measurement with 50 m of SMF fiber used as dispersive fiber in Fig. \ref{setup} is performed with the fiber placed in the arm with the gas cell. In this instance where chirped pulses are sent to the gas cell, the line asymmetry is completely removed. By placing the dispersive fiber in the other arm and allowing short-pulse to travel in the gas cell, the asymmetry arises. It thus suggests that pulse duration and pulse peak intensity influence the asymmetry. This is shown in Fig. \ref{Fano_lines} where the subplots show a single absorption line for different pulses length. By varying the pulse length using a combination of dispersive SMF-28 fiber and/or compensating IDF fiber, the pulse peak power is varied. The top panel of each subplot of Fig. \ref{Fano_lines} shows the 1544.55 absorption line of H$^{12}$CN in color and a fit using a specified theoretical model. The bottom panels show the residuals between the experimental data and the fit. The left column of the figure presents the data with a Voigt fit. As pulse width is reduced and peak power consequently increased, the Voigt fit no longer captures the data. The lines becomes skewed and the residuals becomes dominated by a linear slope as well as a w-shape.   On the other hand, the right column of Fig. \ref{Fano_lines} shows the same experimental data adjusted with the Voigt-Fano model from \cite{SCH18}. The residuals clearly show that this model better captures the absorption lines and its distortion for high peak power as they appear mainly limited by measurement noise.

This Voigt-Fano fit based on Eq. \ref{Fano}  from \cite{SCH18} is a function of the line depth $a$, the $q$ parameter, the gaussian width $\Delta_G$, and the Faddeeva function $w$.

\begin{align}
F = \frac{a}{q^2}\frac{2\sqrt{\ln{2}}}{\Delta_G\sqrt{\pi}} \left\{(q^2-1)\Re(w) - 2q \Im(w) \right\}
\label{Fano}
\end{align}

The $q$ parameter can be expressed as a function of the Fano phase $\phi$ \cite{OTT13}. 

\begin{align}
    q(\phi) = -\cot{\frac{\phi}{2}}
\end{align}

\begin{figure}[!ht]
\centering\includegraphics[width=0.5\textwidth]{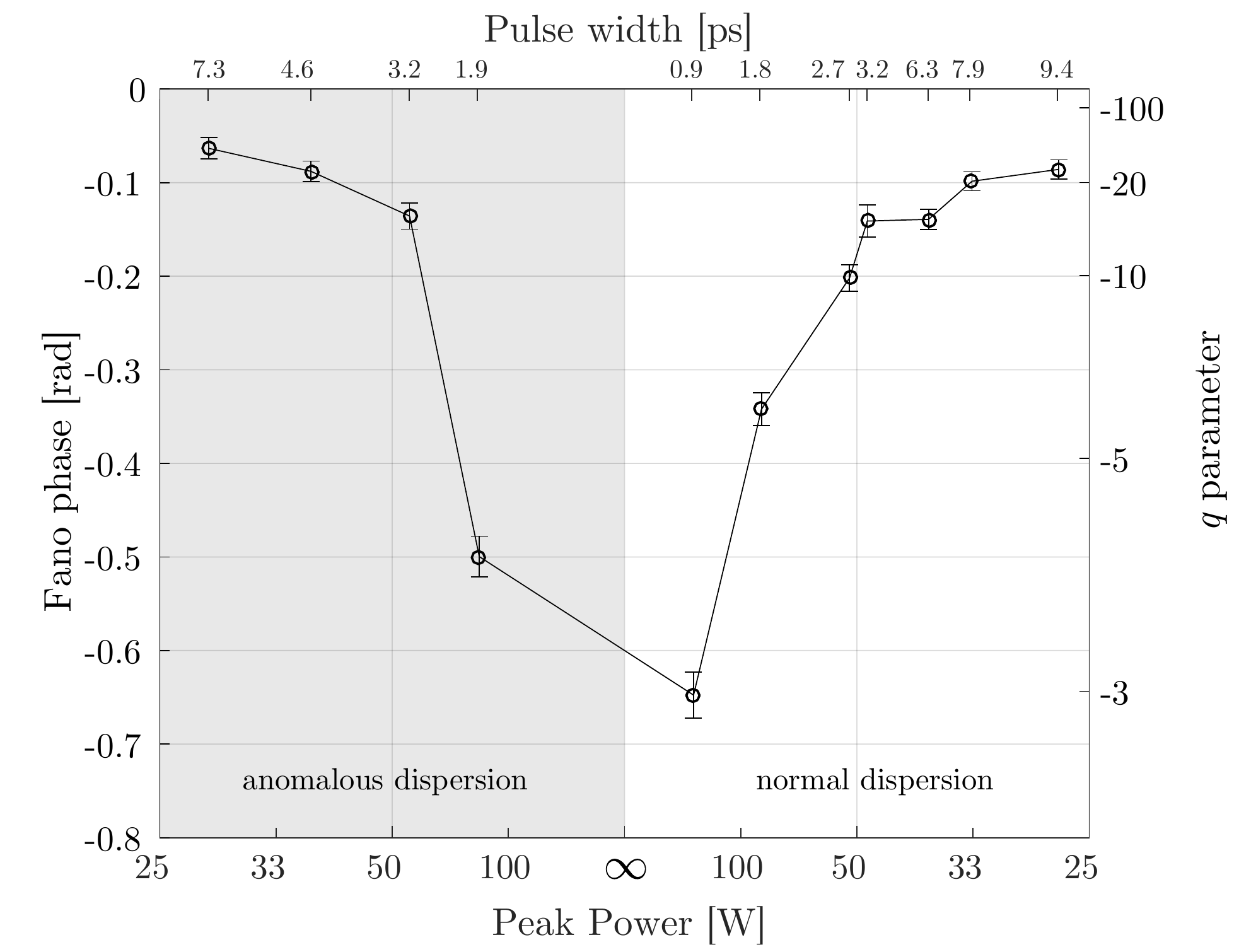}
\caption{Fano phase (left axis) as a function of peak power (bottom axis) and pulse width (top axis). The right axis shows the the $q$ parameter. The left-hand side of the figure (grey area) show the anomalous regime while the right-hand side show the normal dispersion regime.}
\label{Fano_plot}
\end{figure}

The Voigt-Fano fit uses Fano phase, line position, line depth, and Lorentzian and Gaussian width are free parameters. The line position and the Lorentzian width are parameters set in Faddeeva function. Fano phase from fits across different pulse width is plotted against the pulse peak power in Fig. \ref{Fano_plot}. Since the relation between Fano phase and pulse peak power is inverse, the pulse peak power axis has been built to converge in the middle of the figure at infinity. This way, Fano phase from anomalous and normal dispersed pulses converge towards a minimal value at maximum peak power. The pulse width has also been added to the top of the axis and a shaded area has been added to distinguish the anomalous and normal dispersion regime. Moreover, the $q$ parameter is also shown on the right axis. The uncertainty bars are given by the 95\% confidence interval in the fitting procedure.

Fig. \ref{Fano_plot} shows that the Fano phase becomes essentially zero as pulse width increases up to 10 ps. So, it becomes possible to account for the Fano phase with a Voigt fit adjusted by the Fano profile or by ensuring that 30 mW pulses of average power have a 10 ps length. So for a 50 nm bandwidth comb, 10 m of standard single-mode fiber with a dispersion of 20 ps/(nm km) can be used to provide sufficient pulse broadening and avoid any line asymmetry.

\begin{figure}[!ht]
\centering\includegraphics[width=0.5\textwidth]{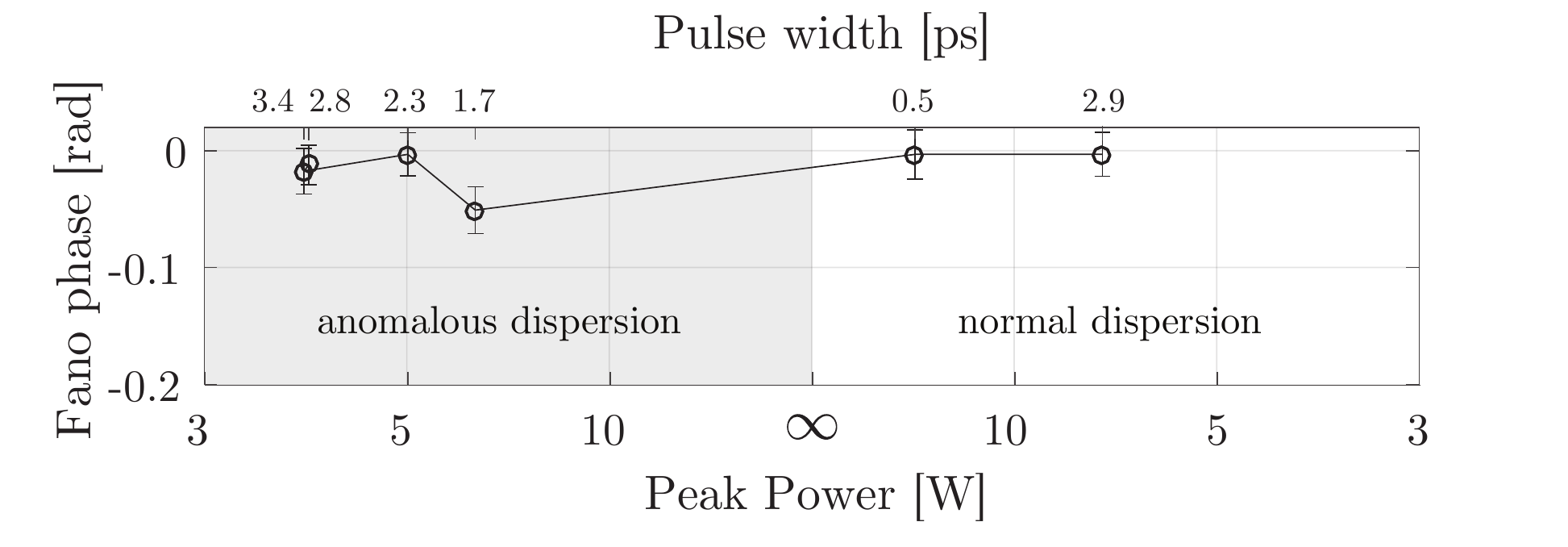}
\caption{Fano phase (left axis) as a function of peak power (bottom axis) and pulse width (top axis). The right axis shows the the $q$ parameter. The left-hand side of the figure (grey area) show the anomalous regime while the right-hand side show the normal dispersion regime.}
\label{Fano_plot2}
\end{figure}

To demonstrate that pulse peak power is indeed the source of Fano coupling, rather than a rapid excitation with deltalike pulses compared to the molecule's response time, the experiment has been repeated with 3 mW interferograms by adding a 10 dB attenuator to each laser output. It is thus expected that absorption line asymmetry will not appear even with sub-picosecond excitation. Fig. \ref{Fano_plot2} shows that Fano phase remains negligible.  

In conclusion, distortion of absorption lines explained by the Fano effect is presented in dual-comb spectroscopy. Pulse peak power in the gas cell is varied to show increased asymmetry in the gas' absorption features. It has been demonstrated by the inability to fit lines with a standard Voigt profile that peak power in the gas cell is a concern and that chirping the optical pulses with standard SMF fiber is sufficient to obtain symmetric absorption features.

\begin{backmatter}
\bmsection{Funding} This work was supported by Natural Sciences and Engineering Research Council of Canada (NSERC).


\bmsection{Disclosures} The authors declare no conflicts of interest.

\bmsection{Data availability} Data underlying the results presented in this paper are not publicly available at this time but may be obtained from the authors upon reasonable request.

\end{backmatter}




\end{document}